\documentclass[twocolumn,prl,showpacs,superscriptaddress]{revtex4-1}
\usepackage{amsmath,amssymb,pgf,graphicx,color,url}
\newcommand{\be}{\begin{equation}}
\newcommand{\ee}{\end{equation}}

\begin{document}
\title{Reconciling Planck results with low redshift astronomical measurements}

\author{Zurab~Berezhiani}
\affiliation{Dipartimento di Fisica e Chimica, Universit\`a di L'Aquila, 67100 Coppito, L'Aquila, Italy} 
\affiliation{INFN, Laboratori Nazionali del Gran Sasso, 67010 Assergi,  L'Aquila, Italy}
\author{A.D.~Dolgov}
\affiliation{Novosibirsk State University, Novosibirsk 90, Russia} 
\affiliation{Dipartimento di Fisica, Universit\`a di Ferrara, 44124 Ferrara, Italy}
\author{I.I. Tkachev}
\affiliation{Institute for Nuclear Research, Russian Academy of Sciences, Moscow, 188300, Russia}
\affiliation{Novosibirsk State University, Novosibirsk 90, Russia}

\begin{abstract}
We show that emerging tension between the direct astronomical measurements at low redshifts  and cosmological parameters deduced from the {\it Planck} measurements of the CMB anisotropies can be alleviated if the dark matter consists of two fractions, stable part being dominant and a smaller unstable fraction. The latter constitutes $\sim 10$ per cent at the recombination epoch if decays by now. 
\end{abstract}
\maketitle

\section{Introduction}

Measurements of the cosmic microwave background (CMB) fluctuations  by  {\it WMAP}~\cite{Hinshaw:2012aka} and {\it Planck}~\cite{Ade:2013zuv}  Collaborations opened a new era in high precision cosmology. These  data are well described by the standard spatially-flat $\Lambda$CDM cosmology with a power low spectrum of adiabatic scalar perturbations, and make a great step towards the precise determination  of cosmological parameters, in particular  of the Hubble constant $H_0 = 100\, h~{\rm km~s}^{-1}~{\rm Mpc}^{-1}$, the density parameters $\Omega_b$ and $\Omega_{dm}$ for the baryon and dark matter fractions, and therefore of the whole matter density $\Omega_m = \Omega_b + \Omega_{dm}= 1-\Omega_\Lambda$. Recently the latest results of the {\it Planck} Collaboration have been published \cite{Planck:2015xua} based on the full mission {\it Planck} data.  They are in excellent agreement with the 2013 data \cite{Ade:2013zuv} but with improved precision.

The {\it Plank} data  imply a rather low value for the  Hubble constant, which is in tension with the direct astronomical measurements of $h$. The {\it Planck} 2015  TT,TE,EE+lowP data, which we take in this Letter as a benchmark, determine the Hubble constant with 1 per cent precision, $h = 0.6727\pm 0.0066$ \cite{Planck:2015xua}.

Direct astronomical measurements of the Hubble constant indicate  larger values. The analysis  \cite{Riess:2011yx} of the Hubble Space Telescope (HST) data based on over 600 cepheids in host galaxies and 8 samples of SNe Ia yields $h = 0.738 \pm 0.024$ including both statistical and systematic errors. Independent analysis of  Carnegie Hubble program \cite{Freedman:2012ny} using the Spitzer Space Telescope data for calibration purposes lead to $h = 0.743 \pm 0.021$. Both these results are discordant with the Planck result at about 2.5$\sigma$ level. Other astronomical estimates also typically imply high values of the Hubble constant. For example, the analysis of the gravitational lensing time delay measurements of the system RXJ1131-1231 implies $h= 0.787 \pm 4.5$  \cite{Suyu:2012aa}.

In addition to $h$, there is tension between other CMB derived observables and their direct low redshift measurements. 
The {\it Planck} results show a tension between the cosmological constraints on $\sigma_8$ and $\Omega_m$ from the CMB \cite{Ade:2013lmv,Ade:2015fva} and from clusters as cosmological probes. 
Cluster data prefer lower values of these observables deviated at more than $2\sigma$ level, see e.g. Refs.~\cite{Vikhlinin:2008ym,Bohringer:2014ooa}.

Recently Baryon Acoustic Oscillations (BAO) in the Ly$\alpha$ forest of BOSS DR11 quasars have been studied at redshift z=2.34~\cite{Font-Ribera:2014wya,Delubac:2014aqe}. The measured position of the BAO peak determines the angular distance, $D_A(z)$ and expansion rate, $H(z)$. Obtained  constraints  imply values of $D_A$ and $H$ that are, respectively, 7\% low and 7\% high compared to the predictions of a flat $\Lambda$CDM cosmological model with the best-fit Planck parameters. The significance of this discrepancy is approximately $ 2.5\sigma$~\cite{Delubac:2014aqe}.

The tension between CMB based determination of several observables by the  {\it Planck} Collaboration and direct low $z$ measurements  is intriguing and deserves attention. The cause of  discrepancy may lie in some calibration errors. On the other hand, it may  hint to a deficiency of the standard  $\Lambda$CDM paradigm. In this paper we show that this discrepancy may be resolved if a certain fraction of dark matter  is unstable. Decaying Dark Matter (DDM) models have been considered previously, see e.g. the most recent Refs \cite{Audren:2014bca,Blackadder:2014wpa}, with the stringent constraint on DDM decay width $\Gamma$. However, in these papers it was assumed that the whole of DM is susceptible to the decay, concluding that the decay time must be larger than 100 Gyr or so. We instead assume that dark matter consists of two fractions, the stable dark matter being dominant while a subdominant unstable part decays between recombination and the present epoch.

\section{Decaying Dark Matter}

 \paragraph{Planck constraints.}
 
To ensure that our model fits the Planck data we accept Planck derived values for all cosmological parameters relevant at recombination. In particular, this means that the sum of initial densities of stable and decaying components of dark matter is fixed, and after formal redshift to the present moment is determined by the  Planck value $\omega_{sdm} + \omega_{ddm} = 0.1198$. 
In our model we vary the initial fraction of decaying component in the cosmological mass density
\be
F \equiv \frac{ \omega_{ddm}}{\omega_{sdm} + \omega_{ddm}}.
\ee
We assume that decay occurs into invisible massless particles and does not produce too many photons.  
Alternatively, one can consider a scenario when dark matter consists of two particle species with masses 
$M+\mu$ and $M-\mu$, and the heavier component decays into the lighter one with emission of invisible 
massless particles. In this case the dark mass fraction disappearing due to decay is equivalent to  $F=\mu/M$. 
Throughout the paper we normalize the width  of the decaying component  $\Gamma$ to km/s/Mpc, i.e. 
in the same units as $H_0$.  
$\Gamma$ is another independent cosmological parameter in our model which we also vary for fitting the data.  
It is bounded from above  by the requirement that the unstable fraction does not decay substantially 
before the last scattering to measurably affect the CMB. Hence, we take $\Gamma < 5000$ in which 
range observed CMB spectra are not altered by decays.

\begin{figure}
\includegraphics[width=0.48\textwidth]{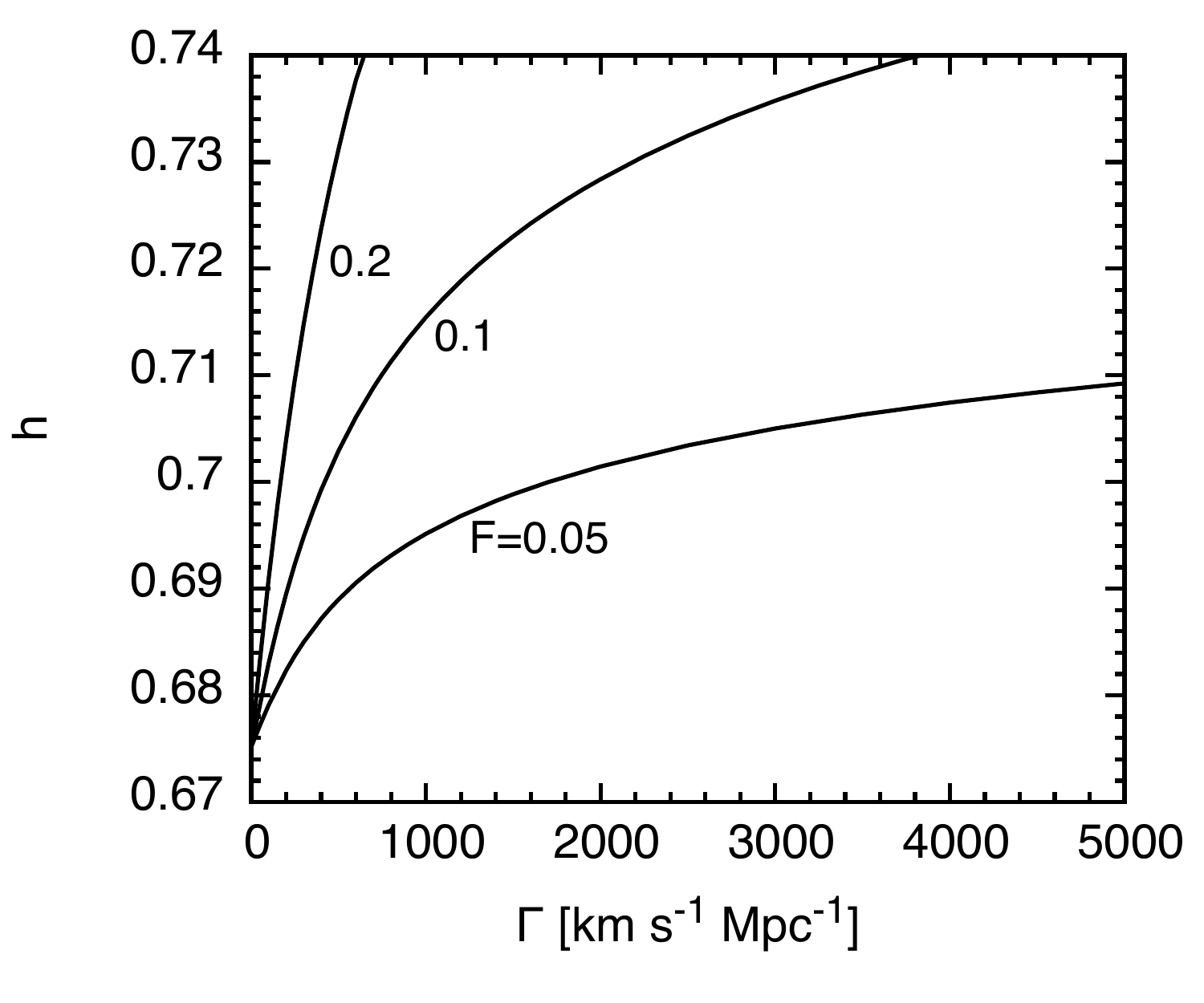}
\caption{
Hubble parameter $h$ as a function of DM decay width $\Gamma$ for different values of the DDM fraction $F$.
}
\label{fig:G_F}
\end{figure}

Furthermore, we require that the angular diameter distance to the last scattering should be the same for all values of parameters, namely we fix the sound horizon angle $100*\theta_s$  to the Planck value $1.04077$. This determines Hubble parameter $h$ as a function of $F$ and $\Gamma$ and guarantees that  derived CMBR spectra in our model are identical (at high $l$) to the best fit Plank spectrum for all values of parameters. Resulting $h$ as a function of  $\Gamma$ is shown in Fig.~\ref{fig:G_F} for different values of $F$.  Let us remark also that for the choice of parameters as in Fig.~\ref{fig:G_F}, 
the age of the Universe 
$t_0 = \frac13 H_0^{-1} \Omega_\Lambda^{-1/2} \ln\big[(1+\Omega_\Lambda^{1/2})/(1-\Omega_\Lambda^{1/2}) \big]$  
remains nearly the same as predicted by {\it Planck}, $t_0 \approx 13.8$~Gyr, 
since increasing of $H_0$ is compensated by  increasing of dark energy fraction $\Omega_\Lambda$.   

Relevant cosmological calculations have been carried out using the CLASS Boltzmann code~\cite{Lesgourgues:2011re,Blas:2011rf}. The parameter space is explored using the Markov Chain Monte-Carlo technique with the Monte Python package~\cite{Audren:2012wb}. We verified that all CMB spectra are identical at $l \agt 40$. At smaller $l$ the spectra somewhat deviate because  the cosmological constant in our model is typically larger as compared to the standard  $\Lambda$CDM (we consider spatially flat Universe only). However, corresponding  changes are smaller than the cosmic variance. Therefore we do not constrain model parameters using low $l$ Planck data and we use supernova data instead.

\begin{figure}
\vspace{0.2cm}
\hspace{-0.5cm}\includegraphics[width=0.50\textwidth]{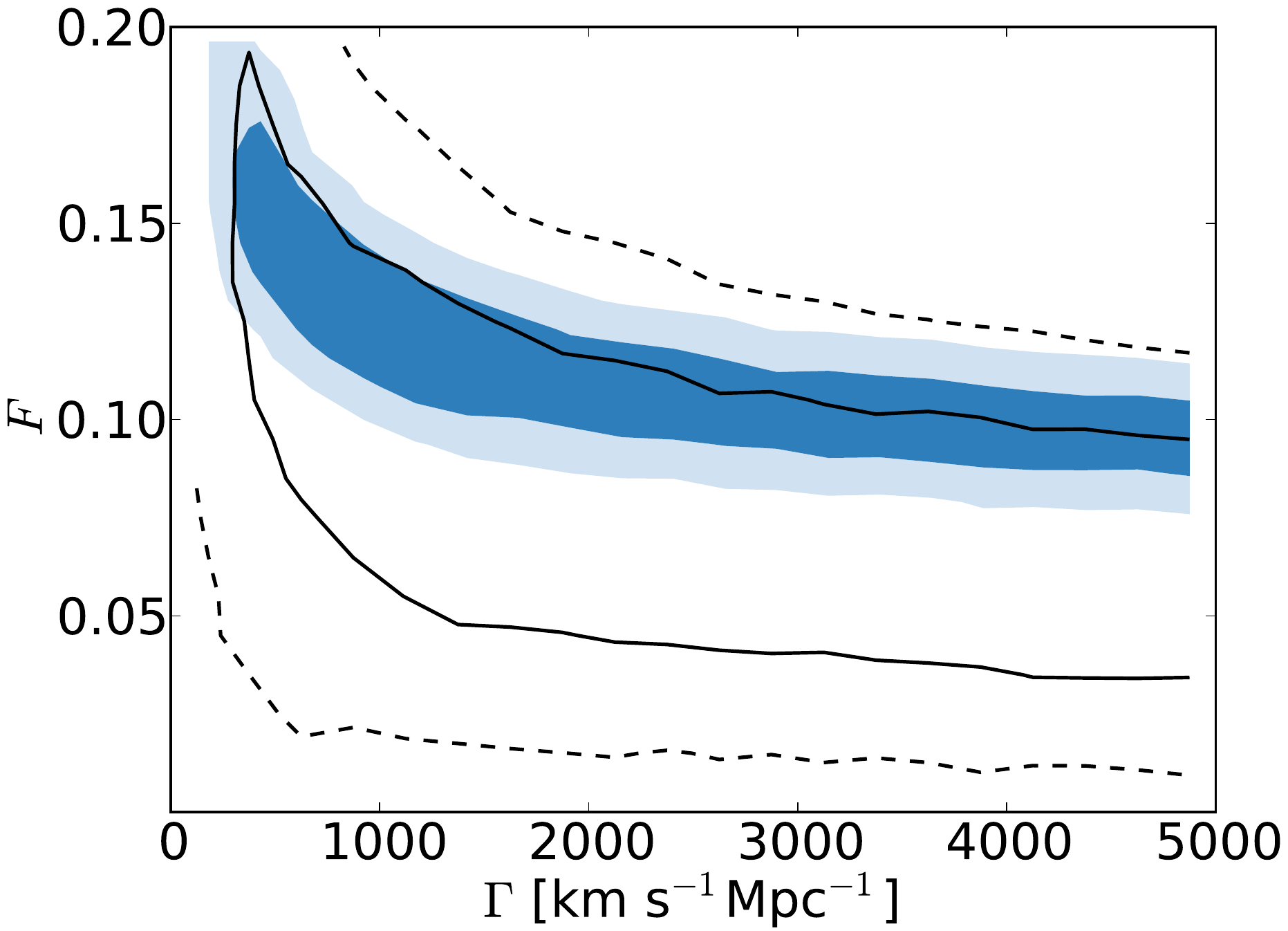}
\caption{One and two sigma likelihood contours for our model parameters. Solid and dashed lines correspond to a dataset consisting of JLA sample of SN Ia and HST measurements of $h$, on top of the best fit Planck model parameters. Addition of Planck cluster data results in much narrower shaded area. } 
\label{fig:L}
\end{figure}

\paragraph{Adding supernova and HST constraints.}

For fitting to  supernovae observations we use  the JLA~\cite{Betoule:2014frx} compilation  composed of 740 SN Ia.  This is the largest data set  to date containing samples from low redshift $z \approx 0.02$ to a large one, $z \approx 1.3$. The data were obtained from the joint analysis of SDSS II and SNLS, improving the analysis by means of a recalibration of light curve fitter SALT2 and in turn reducing possible systematic errors. For "standardization" of SN data the linear model for the distance modulus $\mu$ is employed with four  nuisance parameters in the distance estimates. All necessary data  for the  analysis  were retrieved from~\cite{JLA}. Resulting best fit values for all nuisance parameters in our cosmology do not differ notably from the values quoted in Ref.~\cite{Betoule:2014frx},  derived 
for $\Lambda$CDM.

\begin{figure*}
\includegraphics[width=0.48\textwidth,angle=0]{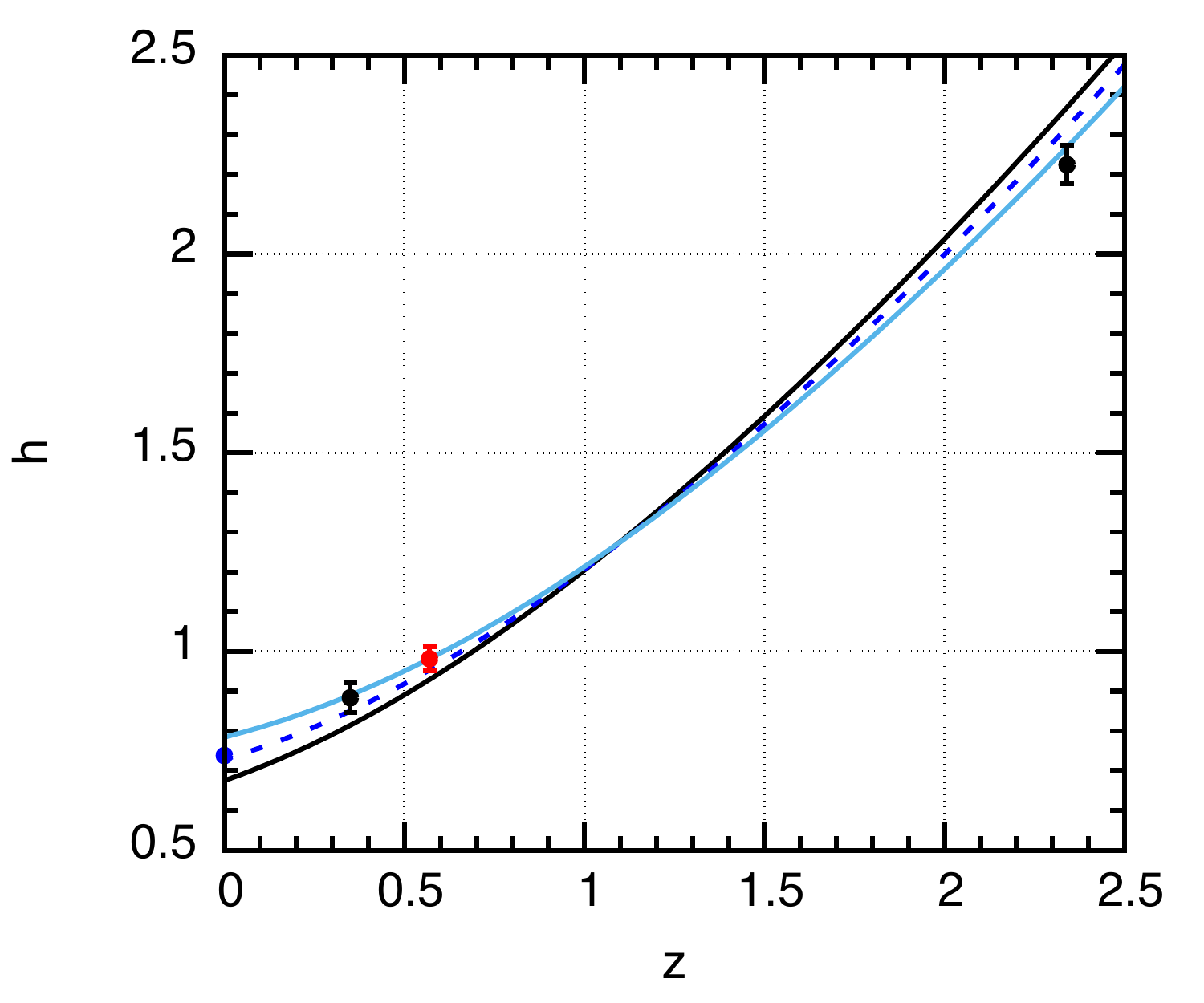}
\label{fig:BAO}
\includegraphics[width=0.48\textwidth,angle=0]{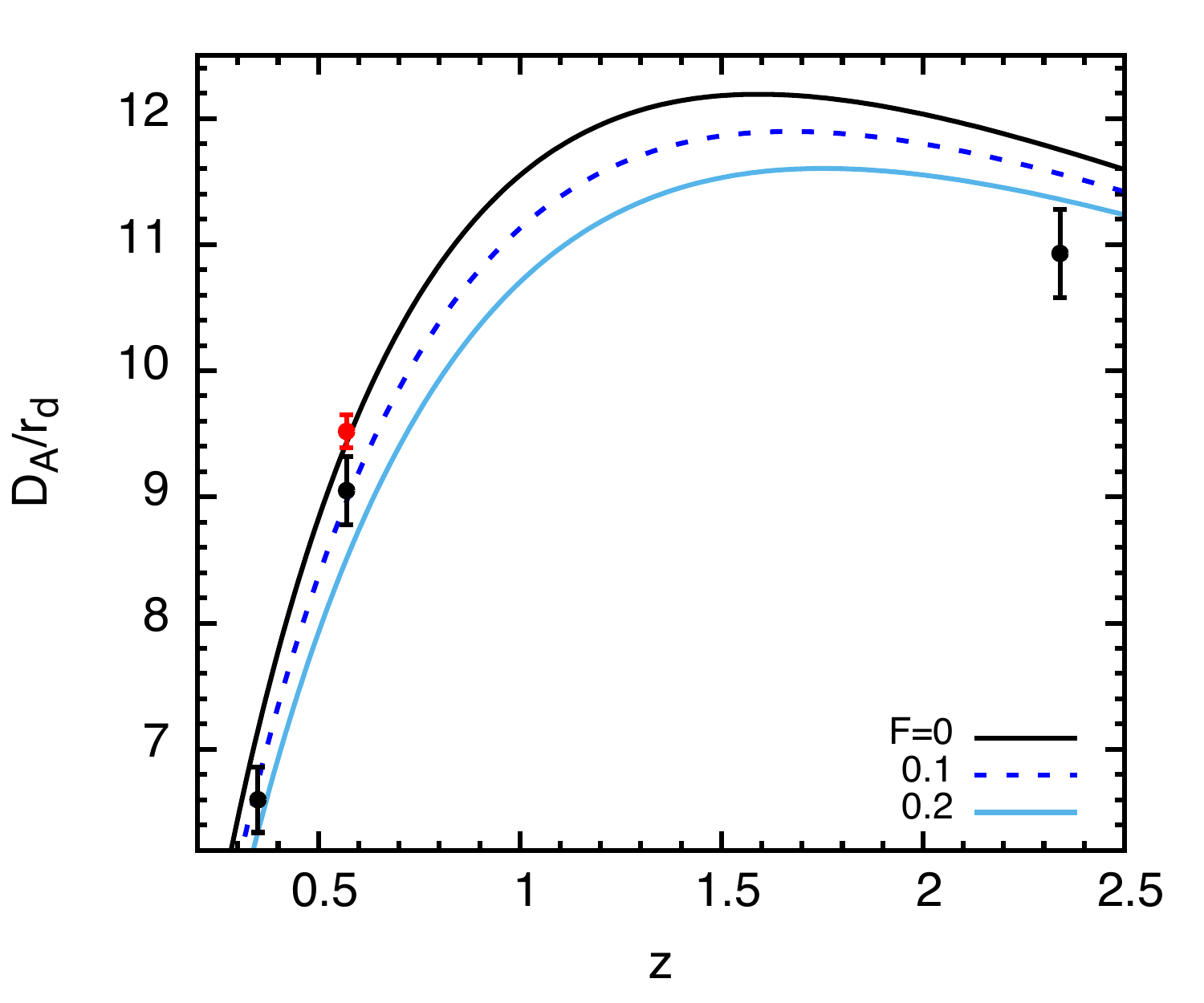}
\caption{
Hubble parameter $h(z)$ (left panel) and angular diameter distance $D_A$ (right panel).  Model curves are presented for fixed $\Gamma = 2000$  and several values of   $F$. Points at non-zero redshift $z$ are the SDSS BAO data. HST measurement at $z=0$ is also shown with  the symbol size  comparable to the errorbars.
}
\label{fig:BAO}
\end{figure*}

We further constrain our model using determination of the Hubble parameter with the HST~\cite{Riess:2011yx}. Resulting one and two sigma likelihood contours in the plane of $\Gamma$ and $F$ are shown in Fig.~\ref{fig:L} by solid and dashed lines. We see that the base $\Lambda$CDM  with $\Gamma = F = 0$ is outside of 2$\sigma$  contours in our model. Derived likelihood for the Hubble parameter corresponds $h = 0.716 \pm 0.02$ at one $\sigma$. Therefore, with a fraction of decaying dark matter the data of Planck on CMBR anisotropies, data on supernova, and HST data all can be reconciled.

\paragraph{DDM and BAO.}
We now turn to the data on Baryon Acoustic Oscillations.
The measurement of the characteristic scale of BAO in the correlation function of different matter distribution tracers provides a powerful tool to probe the cosmic expansion and a convincing method for setting cosmological constraints. The BAO peak in the correlation function at a redshift $z$ appears at the angular separation $\Delta \theta = r_d/(1 + z)D_A(z)$, where $D_A $ is the angular diameter distance and $r_d = r_s(z_d)$ is the sound horizon at the drag redshift, i.e. at the epoch when baryons decoupled from photons. BAO feature also appears at the redshift separation $\Delta z = r_d/D_H$, where $D_H \equiv c/H(z)$. Therefore, measurement of the BAO peak position at some $z$ constrains the combinations of cosmological parameters that determine $D_H/r_d$ and $D_A/r_d$ at that redshift.

Recently independent constraints on $H\, r_d$ and $D_A/r_d$ were obtained using SDSS/BOSS data at $z = 0.35$  \cite{Chuang:2011fy,Xu:2012fw}, z = 0.57 \cite{Kazin:2013rxa,Anderson:2013zyy}, and z = 2.34 \cite{Delubac:2014aqe}. These data are plotted in Fig.~\ref{fig:BAO}. Note that the derived constraints for $H(z)$ and $D_A$ are not independent, the correlation coefficient is 0.5. To avoid cluttering  in displaying results obtained by different authors at the same redshift, in the right panel we plot the results of Refs. \cite{Xu:2012fw,Kazin:2013rxa,Anderson:2013zyy,Delubac:2014aqe} for $D_A/r_d$, while in the left panel the results  of Refs. \cite{Chuang:2011fy,Anderson:2013zyy,Delubac:2014aqe} for $h$ are presented.  Again, the solid line corresponds to the $\Lambda$CDM model with the best fit {\it Planck} measurements. 
Two other models, $F=0.1$ and $F=0.2$ both  with $\Gamma = 2000$, are also shown. 
We see that the data systematically deviate from the  base $\Lambda$CDM. 
Thought at $z<1$ each deviation is about 1$\sigma$ (and therefore this is not considered as a problem) 
they all are in the direction of the DDM models, except for the result of ref.~\cite{Anderson:2013zyy} at $z=0.57$. 

We repeated procedure of likelihood analysis of previous subsection with BAO data added. We used BOSS BAO likelihoods included into Monte Python package~\cite{Audren:2012wb}, latest release  2.1. The result is similar to the one presented in Fig.~\ref{fig:L} but with one and two sigma contours shifted down by about a factor of two. However, the result would clearly depend upon the dataset chosen. In Ref.~\cite{Aubourg:2014yra} DDM model was analyzed with the inclusion of the latest BAO results only, at $z=0.57$ \cite{Anderson:2013zyy} and $z = 2.34$ \cite{Delubac:2014aqe}, with a pessimistic conclusions. In Fig.~\ref{fig:BAO} we can see the origin for this conclusion as well. DDM helps to ease tension at  $z = 2.34$ both for $H(z)$ and 
$D_A$, which are at the 2.5$\sigma$ level compared to the predictions of the base $\Lambda$CDM. However, results of~\cite{Anderson:2013zyy} at $z=0.57$, which are also discrepant at 1$\sigma$ level, behave differently. While DDM is  better fit for $H(z)$, it is not so for $D_A$, the latter is represented by an upper (red) datapoint at $z=0.57$ in the left panel of Fig.~\ref{fig:BAO}.  Overall, DDM does not help much here. As Ref.~\cite{Planck:2015xua} describes, at present it is not clear whether the discrepancy at  $z = 2.34$ is caused by systematics in the Ly$\alpha$ BAO measurements (which are more complex and less mature than galaxy BAO measurements) or it is an indicator of a new physics.

\paragraph{DDM and cluster counts.}

\begin{figure}
\includegraphics[width=0.48\textwidth,angle=0]{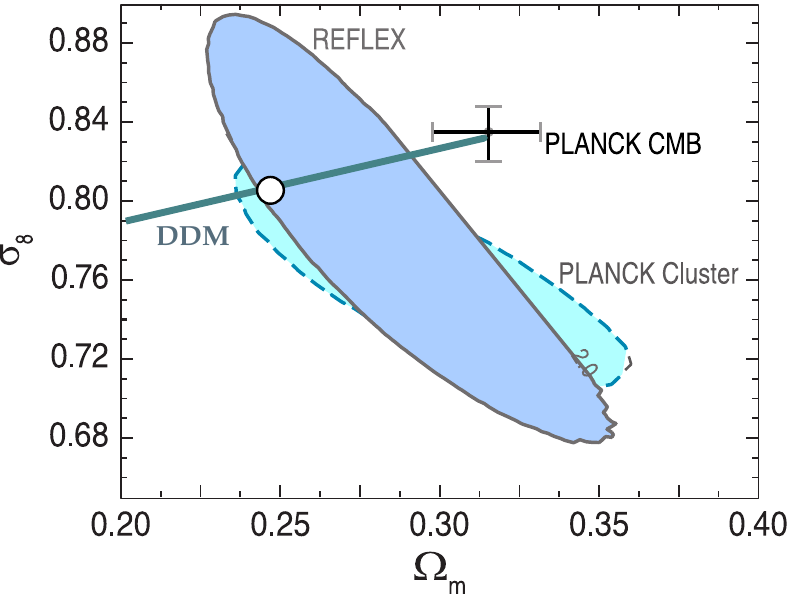}
\caption{
$\Omega_m$ and $\sigma_8$ derived from cluster counts and from CMB.  Line marked DDM shows trend of these parameters when $F$ and $\Gamma$ are varied in our model. White circle represents a model with $F=0.1$ and $\Gamma = 2000$  as an example. 
}
\label{fig:sigma8}
\end{figure}

Decaying Dark Matter model is capable to resolve tension between the base  $\Lambda$CDM model and the cluster data as well  \cite{Aoyama:2014tga}.  This is displayed in Fig.~\ref{fig:sigma8} in the $\sigma_8$ and $\Omega_m$ parameter plane. The base  $\Lambda$CDM corresponds to the error cross marked PLANCK CMB. Shaded areas correspond to the parameter regions allowed (at $2\sigma$) by {\it Planck} cluster data \cite{Ade:2013lmv,Ade:2015fva} and by extended ROSAT-ESO Flux Limited X-ray Galaxy Cluster Survey (REFLEX II) \cite{Bohringer:2014ooa}. We should also note that earlier results obtained in  \cite{Vikhlinin:2008ym}, while in agreement with \cite{Bohringer:2014ooa}, are even father away from the Planck base  $\Lambda$CDM model. In DDM model, when $F$ and $\Gamma$ are varied, $\sigma_8$ and $\Omega_m$ closely follow the line marked DDM in Fig.~\ref{fig:sigma8} and  cross the region  allowed  by the cluster data.  White circle on this line  represents a model with $F=0.1$ and $\Gamma = 2000$. With smaller values of $F$ and/or $\Gamma$  the dot representing a model moves to right, closer to the base  $\Lambda$CDM model.

We have added Planck constraints \cite{Ade:2013lmv} from Sunyaev-Zeldovich cluster counts, $\sigma_8\left(\Omega_m/0.27\right)^{0.3} = 0.78 \pm 0.01$, to our  likelihood analysis (without BAO). Result is shown in Fig.~\ref{fig:L} as shaded area. ($1\sigma$ area continues actually up to $F \approx 0.25$ and $\Gamma \approx 100$ but this is unresolved at the scale of this figure.) Now the  likelihood of base $\Lambda$CDM is vanishingly small compared to a best fit DDM models. However, as Planck collaboration concluded on cluster counts issue~\cite{Ade:2015fva}: it is unclear if this  tension arise from low-level systematics in the astrophysical studies, or represents the first glimpse of something more important. 
We would say again that the hypothesis of decaying dark matter may help to resolve this tension as well. 
In fact, from the joint fit shown in Fig.  ~\ref{fig:L} one can see that the issues of $H_0$  and $\sigma_8$ 
can be resolved with the same parameter values of the DDM model.

\section{Conclusions}

Cosmological parameters deduced from the {\it Planck} measurements of the CMB anisotropies with unprecedented accuracy are at some tension with direct astronomical measurements of various parameters at low redshifts. We have shown   that  {\it Planck}-inspired $\Lambda$CDM cosmology can be reconciled both with HST measurements and with cluster data within the hypotheses of Decaying Dark Matter. Joint fit to {\it Planck}, supernova, HST and Planck cluster data tells that  if the dark matter  decayed between recombination and the present time, then the unstable fraction should be about 10 per cent at the recombination epoch. Situation with the BAO discrepancies is less clear at present and we should wait to see in which direction the intrigue  will develop.\\[2mm]  

\noindent
\paragraph*{Note Added.} 
After our paper has been submitted to arXiv, Ref. \cite{Enqvist:2015ara} appeared  
where DDM was also suggested as a resolution to the possible tension between the CMB and 
weak lensing  determinations of $\sigma_8$. \\[3mm]

{\bf Acknowledgements} 
\vspace{2mm} 

\noindent
The work of Z.B. was supported in part by the MIUR 
grant PRIN  No. 2012CPPYP7 ``Astroparticle Physics" and in part by 
Rustaveli National Science Foundation grant No. DI/8/6-100/12. 
A.D.  and  I.T. acknowledge  support of the Russian Federation Government Grant No. 11.G34.31.0047. 
Numerical part of the work has been done at the cluster of the Theoretical Division of INR RAS.

\end{document}